\documentclass{menu97}    
\newcommand{\ba}{\begin{eqnarray}}
\newcommand{\ea}{\end{eqnarray}}
\usepackage{epsf,rotate}

\begin{document}
    \setlength{\baselineskip}{2.6ex}

\title{Electromagnetic and Strong Decays \\
in a Collective Model of the Nucleon}
\author{A. Leviatan$^a$ and R. Bijker$^b$\\
{\em 
$^a$Racah Institute of Physics, 
The Hebrew University, Jerusalem 91904, Israel\\
$^b$Instituto de Ciencias Nucleares, 
U.N.A.M., A.P. 70-543, 04510 M\'exico, D.F., M\'exico}}
\baselineskip=13pt

\maketitle

\begin{abstract}
\setlength{\baselineskip}{2.6ex}
We present an analysis of electromagnetic elastic form factors, helicity
amplitudes and strong decay widths of non-strange baryon resonances, 
within a collective model of the nucleon. Flavor-breaking and stretching
effects are considered. Deviations from the naive three-constituents 
description are pointed out.
\end{abstract}

\setlength{\baselineskip}{2.6ex}

\section*{}

Regularities in the observed mass spectra of baryons 
({\it e.g.} linear Regge trajectories and parity doubling) suggest 
that a collective type of dynamics may play a role in the structure of 
baryons. 
It is of interest to test the scope and limitation 
of the collective scenario, contrast it with the single-particle type of 
dynamics present in quark potential models, and identify features in
the experimental data which point at the need for additional degrees of 
freedom to supplement the naive effective description with only 
three constituents. In this 
contribution we present a particular collective model of baryons 
\cite{BIL} and report on calculations of electromagnetic \cite{emff} 
and strong \cite{strong} couplings within this framework.

We consider a collective model in which nucleon and delta
resonances are interpreted in terms of rotations and vibrations of a 
Y- shaped string configuration. The underlying shape is that of an 
oblate-top, with a (normalized) distribution of charges and 
magnetization
\ba
g(\beta) &=& \beta^2 \, \mbox{e}^{-\beta/a}/2a^3 ~.
\label{gbeta}
\ea
Here $\beta$ is a radial coordinate along the string and $a$ is a scale 
parameter. The collective wave functions have the form
$\left| \, ^{2S+1}\mbox{dim}\{SU_f(3)\}_J \,
[\mbox{dim}\{SU_{sf}(6)\},L^P]_{(v_1,v_2);K} \, \right>$. 
The spin-flavor part has the usual $SU_{sf}(6)$ classification and
determines the permutation symmetry of the state.
The spatial part is 
characterized by the labels: $(v_1,v_2);K,L^P$, where $(v_1,v_2)$
denotes the vibrations (stretching and bending) of the string; 
$K$ denotes the projection of the
rotational angular momentum $L$ on the body-fixed symmetry-axis and 
$P$ the parity. The spin $S$ and $L$ are coupled to total angular momentum
$J$. In this notation the nucleon and the delta ground state
wave functions are given by
$\left| \, ^{2}8_{1/2} \, [56,0^+]_{(0,0);0} \, \right> $ and
$\left| \, ^{4}10_{3/2} \, [56,0^+]_{(0,0);0} \, \right>$ respectively.

A collective model analysis \cite{BIL,CONF} of the mass spectrum 
produced a fit for 3* and 4* nonstrange resonances, of comparable quality
to that of non-relativistic \cite{NRQM} and relativized \cite{RQM} 
quark potential models.
This shows that masses alone are not sufficient to distinguish between 
single-particle and collective forms of dynamics and one has to examine
other observables which are more sensitive to the structure of 
wave-functions, such as electromagnetic and strong couplings.
The electromagnetic (strong) transition operators are assumed to involve 
the absorption or emission of a photon (elementary meson) 
from a single constituent. The collective form factors are obtained
by folding the matrix elements of these operators
with the probability distribution of Eq. (\ref{gbeta}).
In Ref. \cite{BIL} these form factors are evaluated algebraically and
closed expressions are derived in the limit of large model space.
The ansatz of Eq.~(\ref{gbeta}) for the probability distribution
is made to obtain the dipole form for the elastic form factor.
The same distribution is used to calculate inelastic form factors 
connecting other final states. 
All collective form factors are found \cite{emff} 
to drop as powers of $Q^2$. 
This property is well-known experimentally and is in contrast with harmonic 
oscillator based quark models in which all 
\noindent\begin{tabular}{p{5.5cm}p{5.5cm}}
\epsfxsize 5.5cm
\rotate{\rotate{\rotate{\epsffile{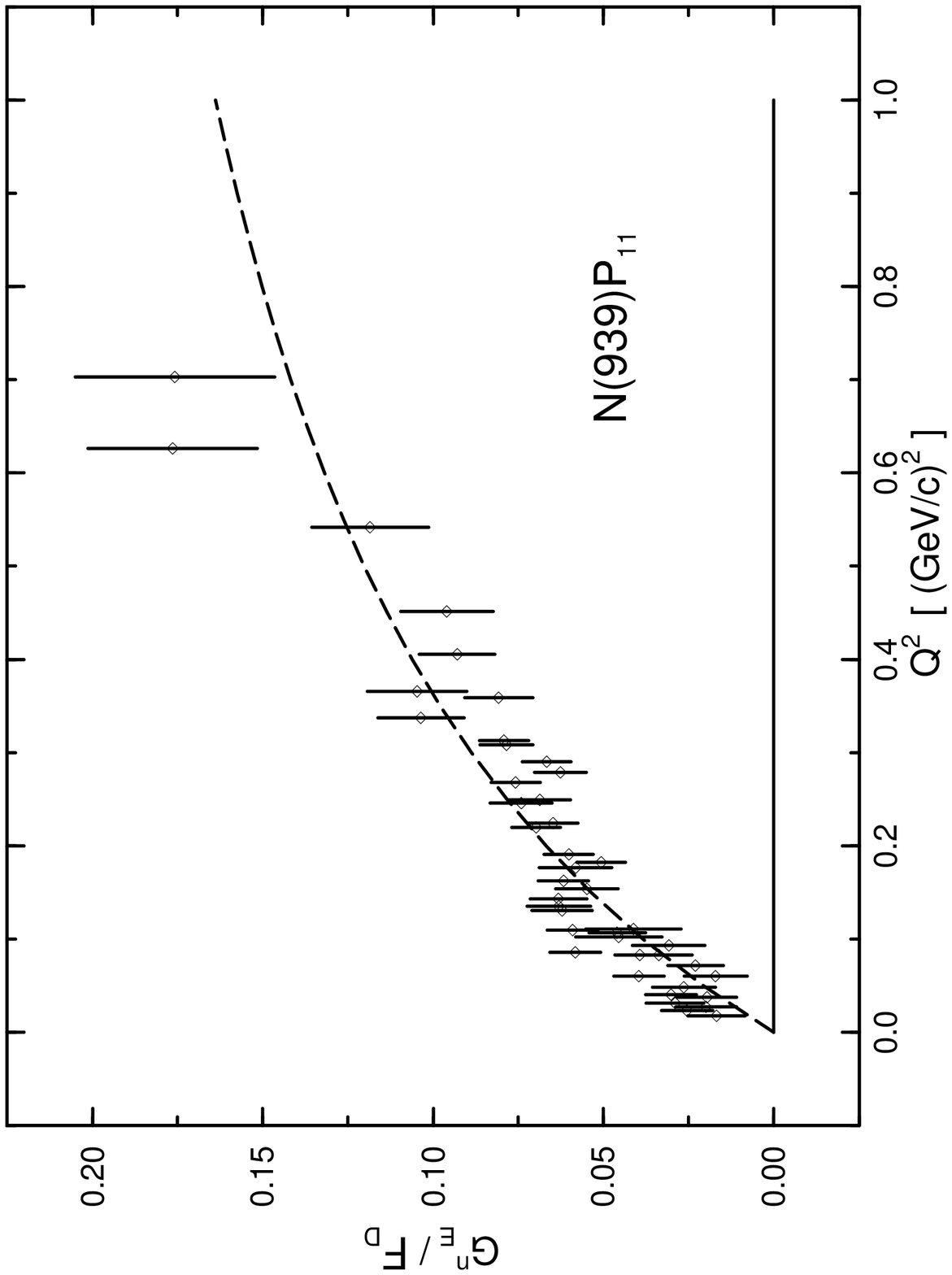}}}}
&
\epsfxsize 5.5cm
\rotate{\rotate{\rotate{\epsffile{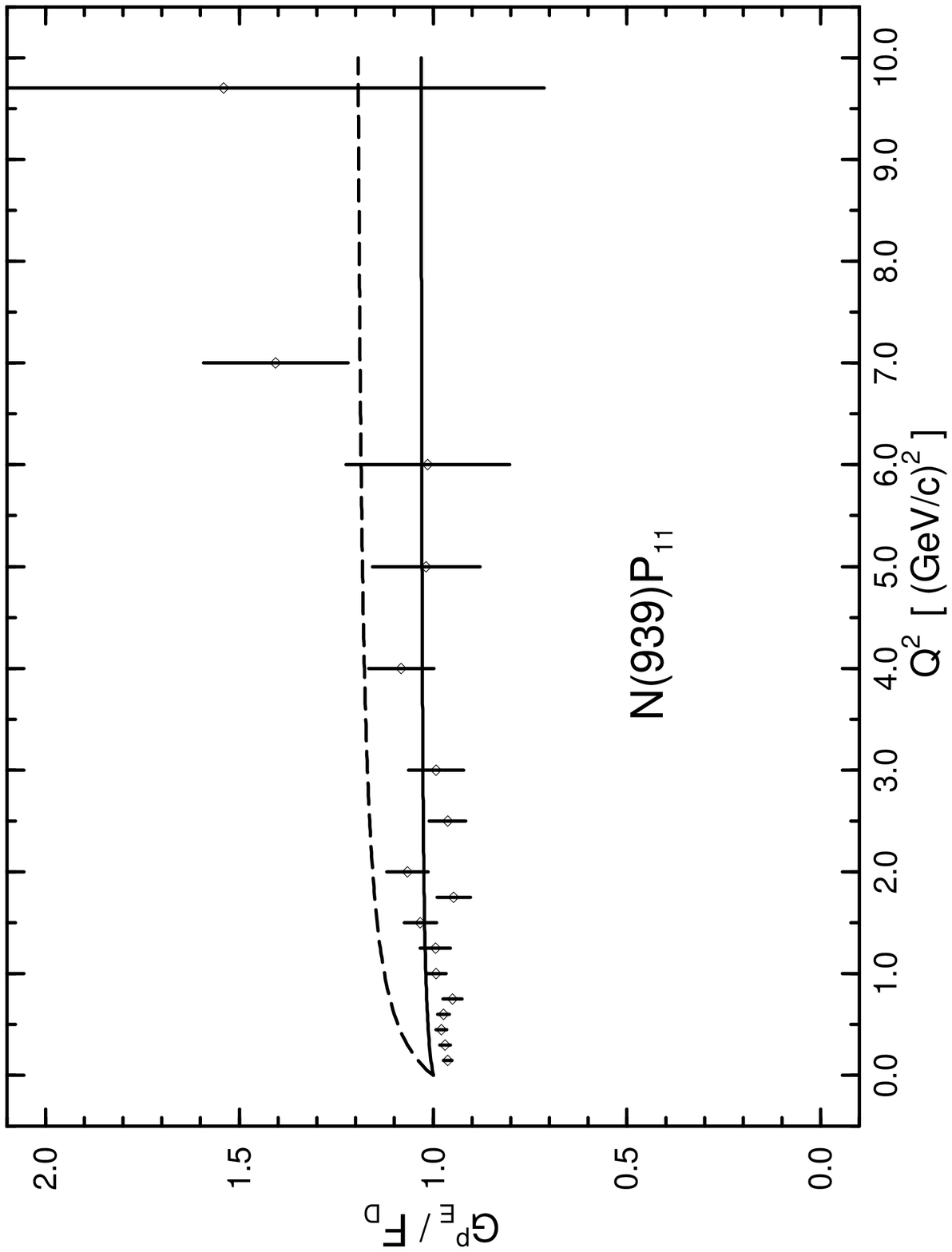}}}}\\
\epsfxsize 5.5cm
\rotate{\rotate{\rotate{\epsffile{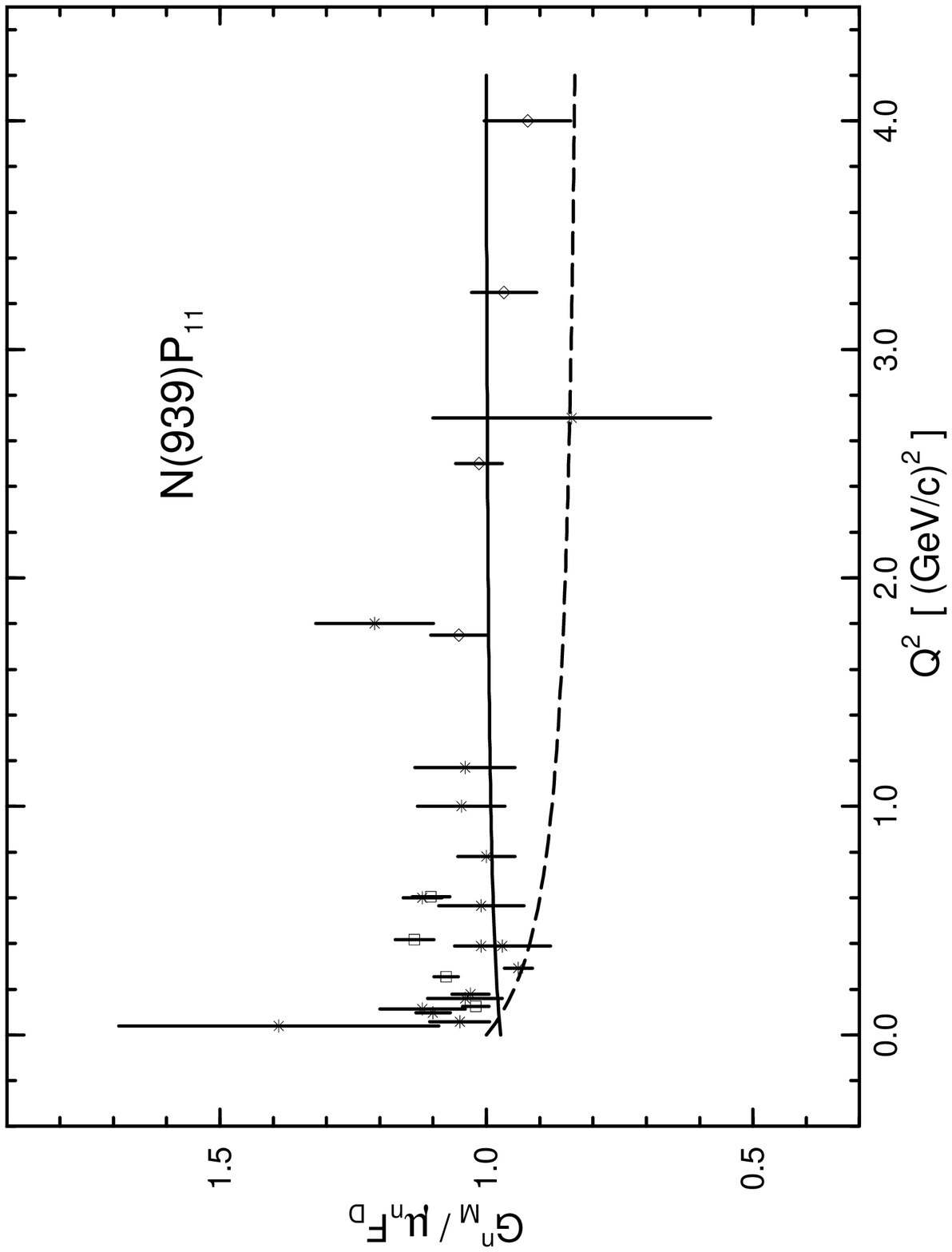}}}}
&
\epsfxsize 5.5cm
\rotate{\rotate{\rotate{\epsffile{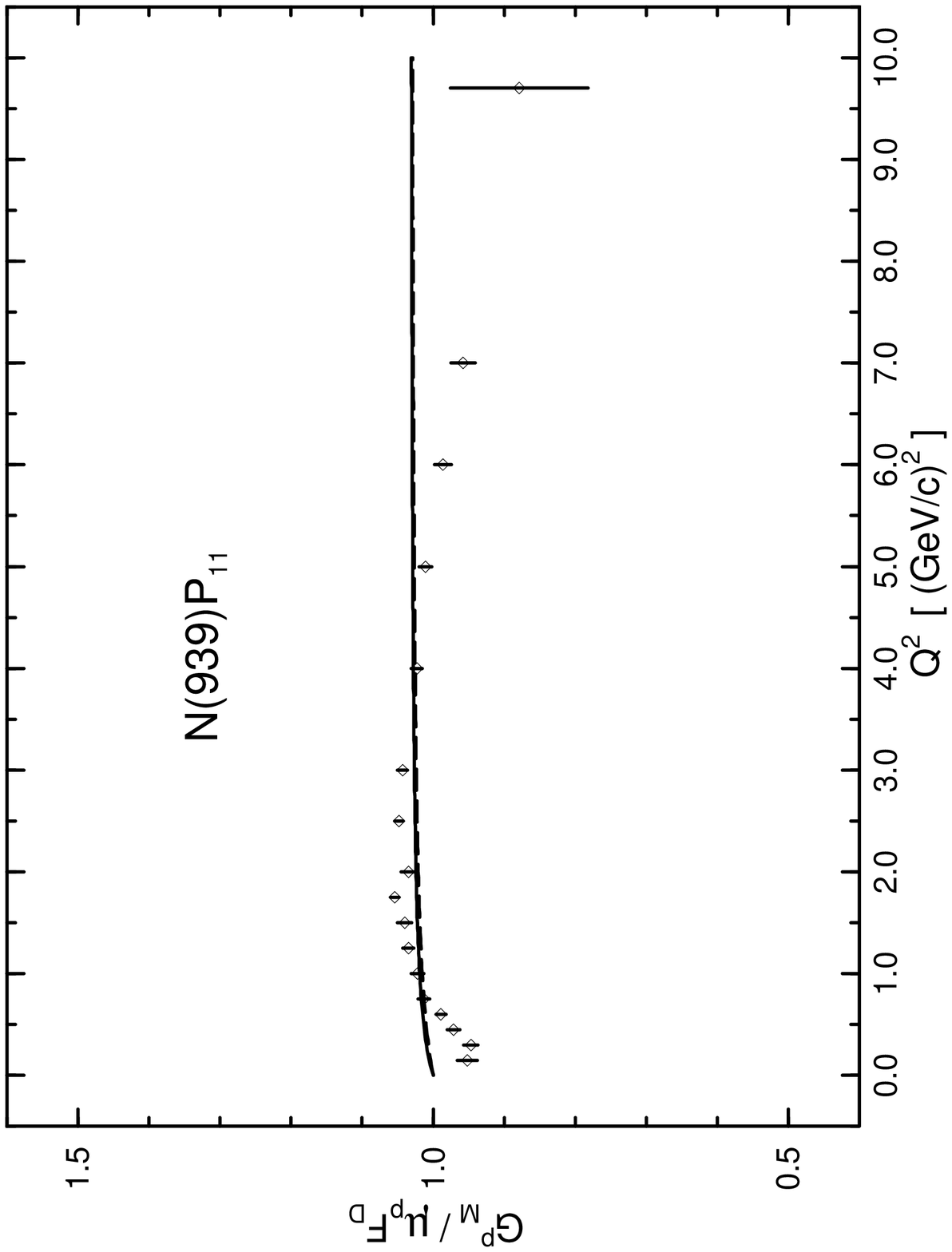}}}}
\end{tabular}
\vspace{3pt}
{\small \setlength{\baselineskip}{2.6ex} Fig.~1. 
Neutron and proton electric ($G_E^n$, $G_E^p$) and magnetic 
($G_M^n/\mu_n$, $G_M^p/\mu_p$) form factors divided\\ 
by $F_D=1/(1+Q^2/0.71)^2$. Dashed (solid) lines 
correspond to a calculation with (without) flavor breaking.}
\vspace{8pt}

\noindent
form factors fall off exponentially.
Within an effective model with three-constituents,
in order to have a nonvanishing neutron electric form factor,
as experimentally observed, one must break $SU_{sf}(6)$. 
We study this breaking by assuming a flavor-dependent distribution
\ba
g_u(\beta) &=& \beta^2 \, \mbox{e}^{-\beta/a_u} /2a_u^3 ~,
\nonumber\\
g_d(\beta) &=& \beta^2 \, \mbox{e}^{-\beta/a_d} /2a_d^3 ~.
\label{gugd}
\ea
The scale parameters $a_u$ and $a_d$ in Eq. (\ref{gugd})
and the scale quark magnetic moments $\mu_u,\mu_d$ 
are determined from a simultaneous fit to the
proton and neutron charge radii, and to the proton and neutron electric 
and magnetic form factors. For the calculations in which the $SU_{sf}(6)$ 
symmetry is satisfied this procedure yields $a_u=a_d=a=0.232$ fm and
$\mu_u=\mu_d=\mu_p= 2.793$ $\mu_N$ ($=0.127$ GeV$^{-1}$).
When $SU_{sf}(6)$ symmetry is broken
we find $a_u=0.230$ fm, $a_d=0.257$ fm, 
$\mu_u=2.777\mu_{N}$ 
($= 0.126$ GeV$^{-1}$) and $\mu_d=2.915\mu_{N}$ ($=0.133$ GeV$^{-1}$).

Fig.~1 shows the electric and magnetic
form factors of the proton and the neutron.
We see that while the breaking of spin-flavor symmetry can
account for the non-zero value of $G_E^n$ and gives a good description
of the data, it worsens the fit to the proton electric and neutron
magnetic form factors. There are also noticeable discrepancies at 
the low-$Q^2$ region $0\leq Q^2\leq 1$ (GeV/c)$^2$. This suggests that 
other contributions, such as coupling to the meson cloud \cite{vmd} 
\noindent\begin{tabular}{p{5.5cm}p{5.5cm}}
\epsfxsize 5.5cm
\rotate{\rotate{\rotate{\epsffile{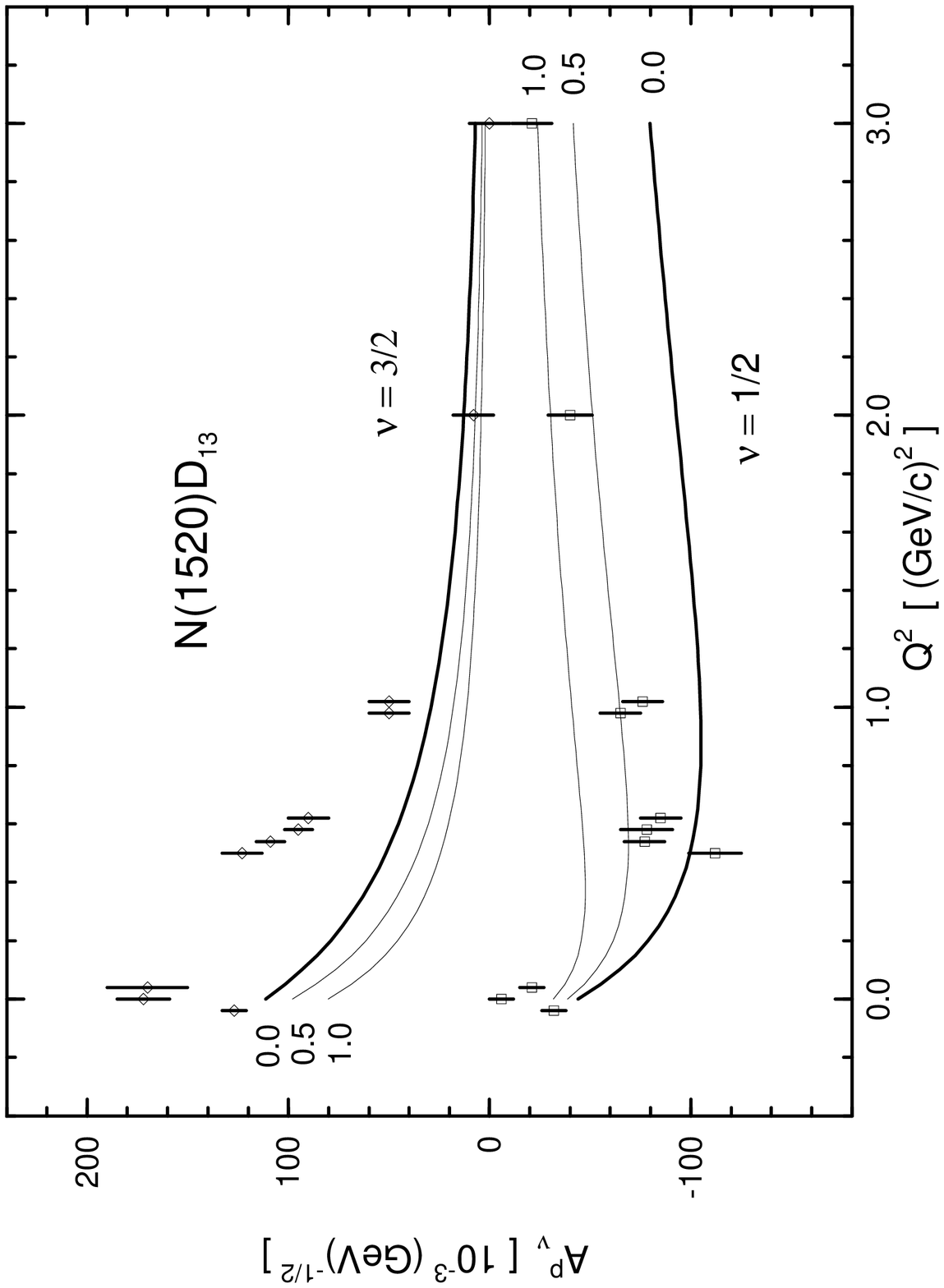}}}}
&
\epsfxsize 5.5cm
\rotate{\rotate{\rotate{\epsffile{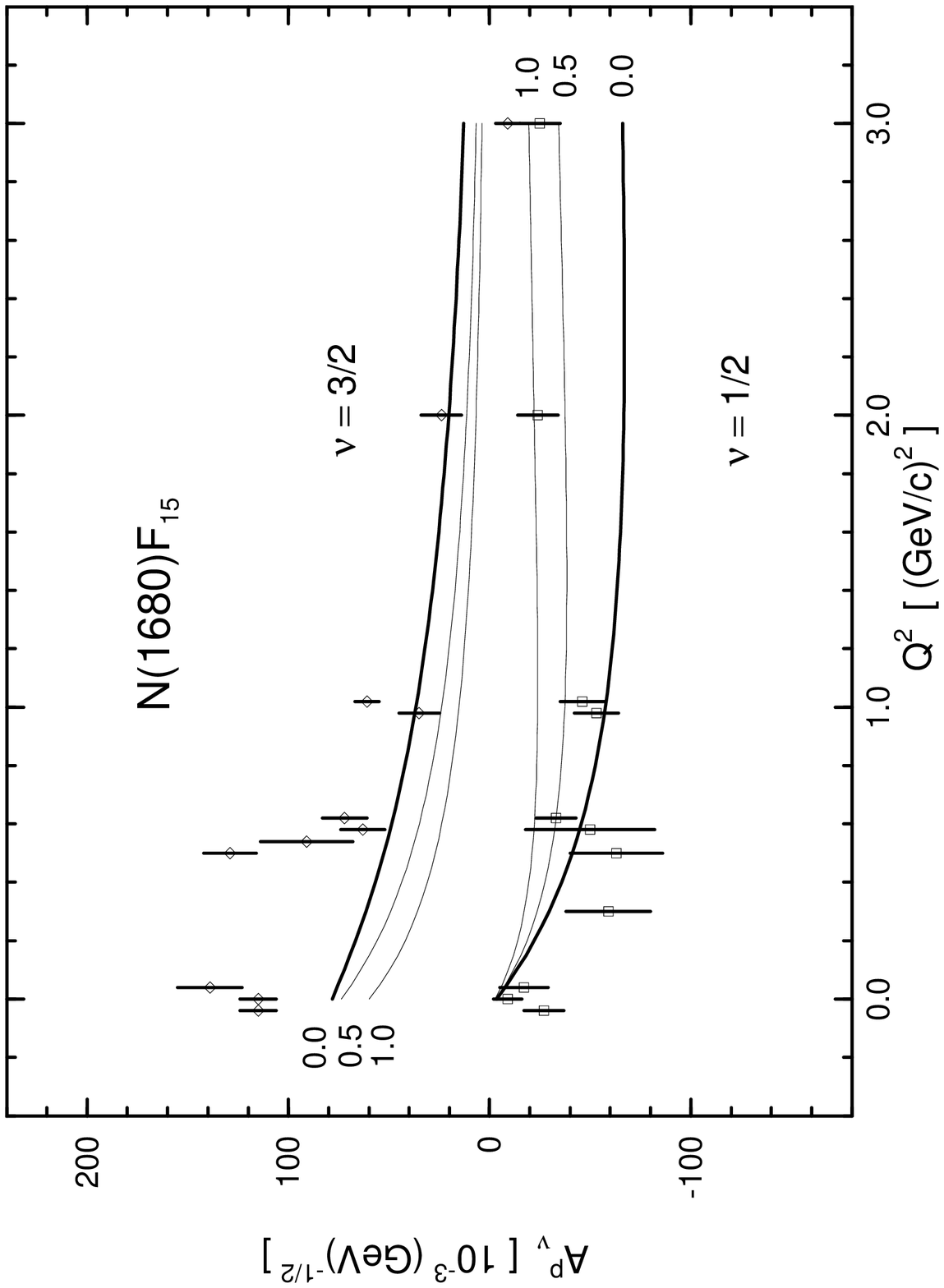}}}}
\end{tabular}
\vspace{3pt}
{\small \setlength{\baselineskip}{2.6ex} Fig.~2. 
Effect of hadron swelling for excitation of
$N(1520)D_{13}$ and $N(1680)F_{15}$. 
The curves are labelled by the stretching parameter $\xi$
of Eq.~(\ref{stretch}).}
\vspace{8pt}

\noindent
contribute in this range of $Q^2$. 
This conclusion ({\it i.e.} worsening the proton form factors)
applies also to the other mechanisms of spin-flavor symmetry
breaking, such as that induced by a hyperfine interaction \cite{iks}. 

The effect of spin-flavor breaking on helicity amplitudes $A_{\nu}$ 
$(\nu=1/2,3/2)$ is rather small.
Only in those cases in which the amplitude with $SU_{sf}(6)$
symmetry is zero, the effect is of some relevance.
Such is the case with proton helicity amplitudes for the
$^{4}8_{J}[70,L^{P}]$ multiplet
({\it e.g.} the $L^{P}=1^{-}$ resonances $N(1675)D_{15}$ and 
$N(1700)D_{13}$) and with neutron $\nu=3/2$ 
amplitudes for the $^{2}8_{J}[56,L^{P}]$ multiplet
({\it e.g.} the $L^{P}=2^{+}$ resonance $N(1680)F_{15}$).
In a string-like model of hadrons one expects \cite{johnsbars} 
on the basis of QCD that strings will elongate (hadrons swell)
as their energy increases. This effect can be
included in the present analysis by making the scale parameters
of the strings energy- dependent
\ba
a &=& a_0\Bigl ( 1 + \xi\,{W-M\over M}\Bigr ) ~. \label{stretch}
\ea
Here $M$ is the nucleon mass, $W$ the resonance mass and $\xi$ 
the stretchability parameter of the string.
Fig.~2 shows that the effect
of stretching on the helicity amplitudes for 
$N(1520)D_{13}$ and $N(1680)F_{15}$ is rather large
(especially if one takes the value $\xi\approx 1$ which is suggested
by QCD arguments \cite{johnsbars} and the Regge behavior of nucleon
resonances). The data show a clear indication that the form
factors are dropping faster than expected on the basis of the dipole
form. 

In addition to electromagnetic couplings, strong decays of baryons
provide an important, complementary, tool to study their structure.
In the algebraic method the widths can be obtained in closed form
which allows us to do a straightforward
and systematic analysis of the experimental data.
\noindent
\begin{table}
\centering
\caption[]{\small
$N \pi$ and $N \eta$ decay widths of (3* and 4*)
nucleon resonances in MeV. 
The experimental values are taken from \cite{PDG}.
\normalsize}
\label{npi}
\vspace{15pt}
\begin{tabular}{lcccccc}
\hline
& & & & & & \\
State & Mass & Resonance & \multicolumn{2}{c}{$\Gamma(N \pi)$}
& \multicolumn{2}{c}{$\Gamma(N \eta)$}\\
& & & th & exp & th & exp \\
& & & & & & \\
\hline
& & & & & & \\
$S_{11}$ & $N(1535)$ & $^{2}8_{1/2}[70,1^-]_{(0,0);1}$
&  $85$ & $79 \pm 38$ &  $0.1$ & $74 \pm 39$ \\
$S_{11}$ & $N(1650)$ & $^{4}8_{1/2}[70,1^-]_{(0,0);1}$
&  $35$ & $130 \pm 27$ &  $8$   & $11 \pm 6$ \\
$P_{13}$ & $N(1720)$ & $^{2}8_{3/2}[56,2^+]_{(0,0);0}$
&  $31$ & $22 \pm 11$ &  $0.2$ & \\
$D_{13}$ & $N(1520)$ & $^{2}8_{3/2}[70,1^-]_{(0,0);1}$
& $115$ & $67 \pm 9$ &  $0.6$ & \\
$D_{13}$ & $N(1700)$ & $^{4}8_{3/2}[70,1^-]_{(0,0);1}$
&   $5$ & $10 \pm 7$ &  $4$   & \\
$D_{15}$ & $N(1675)$ & $^{4}8_{5/2}[70,1^-]_{(0,0);1}$
&  $31$ & $72 \pm 12$ & $17$   & \\
$F_{15}$ & $N(1680)$ & $^{2}8_{5/2}[56,2^+]_{(0,0);0}$
&  $41$ & $84 \pm 9$ &  $0.5$ & \\
$G_{17}$ & $N(2190)$ & $^{2}8_{7/2}[70,3^-]_{(0,0);1}$
&  $34$ & $67 \pm 27$ & $11$   & \\
$G_{19}$ & $N(2250)$ & $^{4}8_{9/2}[70,3^-]_{(0,0);1}$
&  $7$ & $38 \pm 21$ &  $9$   & \\
$H_{19}$ & $N(2220)$ & $^{2}8_{9/2}[56,4^+]_{(0,0);0}$
&  $15$ & $65 \pm 28$ &  $0.7$ & \\
$I_{1,11}$ & $N(2600)$ & $^{2}8_{11/2}[70,5^-]_{(0,0);1}$
&  $9$ & $49 \pm 20$ & $3$ & \\
& & & & & & \\
\hline
\end{tabular}
\end{table}
The calculated values depend on two parameters 
determined from a least square fit to the $N \pi$ partial widths
(which are relatively well known) 
with the exclusion of the $S_{11}$ resonances. 
These parameters are then used to calculate the decay channels 
($N \pi$, $N \eta$, $\Delta \pi$, $\Delta \eta$) of {\em all} resonances. 
The calculation of decay widths of nucleon resonances
into the $N \pi$ channel is found to be in fair agreement with 
experiment (see Table~1). The same holds for the $\Delta \pi$
channel \cite{strong}. These results are to a large
extent a consequence of spin-flavor symmetry. 
There does not seem to be anything unusual in the decays into
$\pi$ and our analysis confirms the results of previous 
analyses \cite{KI,CR}. Contrary to the decays into $\pi$,
the decay widths into $\eta$ have some unusual
properties. The calculation gives systematically small values
for these widths (see Table~1). This is due to a combination of 
phase space factors and the structure of the transition operator. 
In contrast, the 1996 PDG compilation \cite{PDG} 
assigns a large $\eta$ width to $N(1535)S_{11}$ and a small but 
non-zero $\eta$ width to $N(1650)S_{11}$.
The results of our analysis suggest that the large $\eta$ width for the
$N(1535)S_{11}$ is not due to a conventional $q^3$ state. One possible
explanation is the presence of another state in the same mass region,
{\it e.g.} a quasi-bound meson-baryon $S$ wave resonance just below
or above threshold, for example $N\eta$, $K\Sigma$ or $K\Lambda$ 
\cite{Kaiser}. Another possibility is an exotic configuration of four
quarks and one antiquark ($q^{4}\bar{q}$).

The results reported in this article are based on work done
in collaboration with F. Iachello (Yale).
The work is supported in part by grant No. 94-00059 from the United 
States-Israel Binational Science Foundation (BSF), Jerusalem, Israel 
(A.L.) and by DGAPA-UNAM under project IN101997 (R.B.).

\end{document}